\documentclass[conference]{IEEEtran}
\IEEEoverridecommandlockouts

\usepackage{cite}
\usepackage{amsmath, amssymb, amsfonts}

\usepackage{algorithm}
\usepackage{algorithmic}

\usepackage{graphicx}
\usepackage{caption}
\usepackage{subcaption}
\usepackage{booktabs}
\usepackage{multirow}
\usepackage{makecell}
\usepackage{stfloats}

\usepackage{textcomp}
\usepackage{xcolor}
\usepackage{colortbl}

\usepackage{url}
\usepackage{tcolorbox}
\usepackage{enumitem}
\usepackage{setspace}
\usepackage{pifont}

\begin{document}
\bibliographystyle{unsrt}
 % Replace 'references' with the name of your .bib file
 
\title{Towards Visualizing Electronic Medical Records via Natural Language Queries}

\author{\IEEEauthorblockN{
   Haodi Zhang\IEEEauthorrefmark{1}\IEEEauthorrefmark{2},
   Siqi Ning\IEEEauthorrefmark{1},
   Qiyong Zheng\IEEEauthorrefmark{3},
   Jinyin Nie\IEEEauthorrefmark{1},
   Liangjie Zhang\IEEEauthorrefmark{1},
   Weicheng Wang\IEEEauthorrefmark{5},
   Yuanfeng Song\IEEEauthorrefmark{4}}
\IEEEauthorblockA{
\IEEEauthorrefmark{1}College of Computer and Software Engineering, Shenzhen University, Shenzhen, China \\
\IEEEauthorrefmark{2}Hong Kong University of Science and Technology (GZ), Guangzhou, China, \\
\IEEEauthorrefmark{3}School of Mathematical Sciences, Shenzhen University, Shenzhen, China, \\
\IEEEauthorrefmark{4}AI Group, WeBank Co., Ltd, Shenzhen, China\\
\IEEEauthorrefmark{5}Hong Kong University of Science and Technology, Hong Kong, China \\
}
}

\makeatletter

\def\ps@IEEEtitlepagestyle{%
 \def\@oddfoot{}%
 \def\@evenfoot{}%
}

\maketitle

\begin{abstract}
Electronic medical records (EMRs) contain essential data for patient care and clinical research. With the diversity of structured and unstructured data in EHR, data visualization is an invaluable tool for managing and explaining these complexities. However, the scarcity of relevant medical visualization data and the high cost of manual annotation required to develop such datasets pose significant challenges to advancing medical visualization techniques. To address this issue, we propose an innovative approach using large language models (LLMs) for generating visualization data without labor-intensive manual annotation. We introduce a new pipeline for building text-to-visualization benchmarks suitable for EMRs, enabling users to visualize EMR statistics through natural language queries (NLQs). The dataset presented in this paper primarily consists of paired text medical records, NLQs, and corresponding visualizations, forming the first large-scale text-to-visual dataset for electronic medical record information called MedicalVis with 35,374 examples.
Additionally, we introduce an LLM-based approach called MedCodeT5, showcasing its viability in generating EMR visualizations from NLQs, outperforming various strong text-to-visualization baselines.
Our work facilitates standardized evaluation of EMR visualization methods while providing researchers with tools to advance this influential field of application. In a nutshell, this study and dataset have the potential to promote advancements in eliciting medical insights through visualization.

\end{abstract}


\section{Introduction}

Electronic medical records (EMRs) \cite{williams2008role} have become a vital source of patient information for healthcare providers. EMRs contain a comprehensive history of a patient’s health data, including diagnoses, medications, lab test results, medical imaging, doctor’s notes, and more. This structured and unstructured data offers invaluable insights for improving clinical decisions and medical research. However, efficiently analyzing EMR data remains an open challenge. The free-form nature of EMR entries such as clinical notes, and the diversity of data types make it challenging to extract key information and trends. Although text-to-SQL methods in the medical field have made some progress, such as the development of MIMICSQL \cite{wang2020text} and EHRSQL \cite{lee2022ehrsql}, the data retrieved by structured query language (SQL) statements still lack intuitive clarity, necessitating further analysis and exploration. 

\begin{figure}[th!]
 \centering
\includegraphics[width=0.48\textwidth]{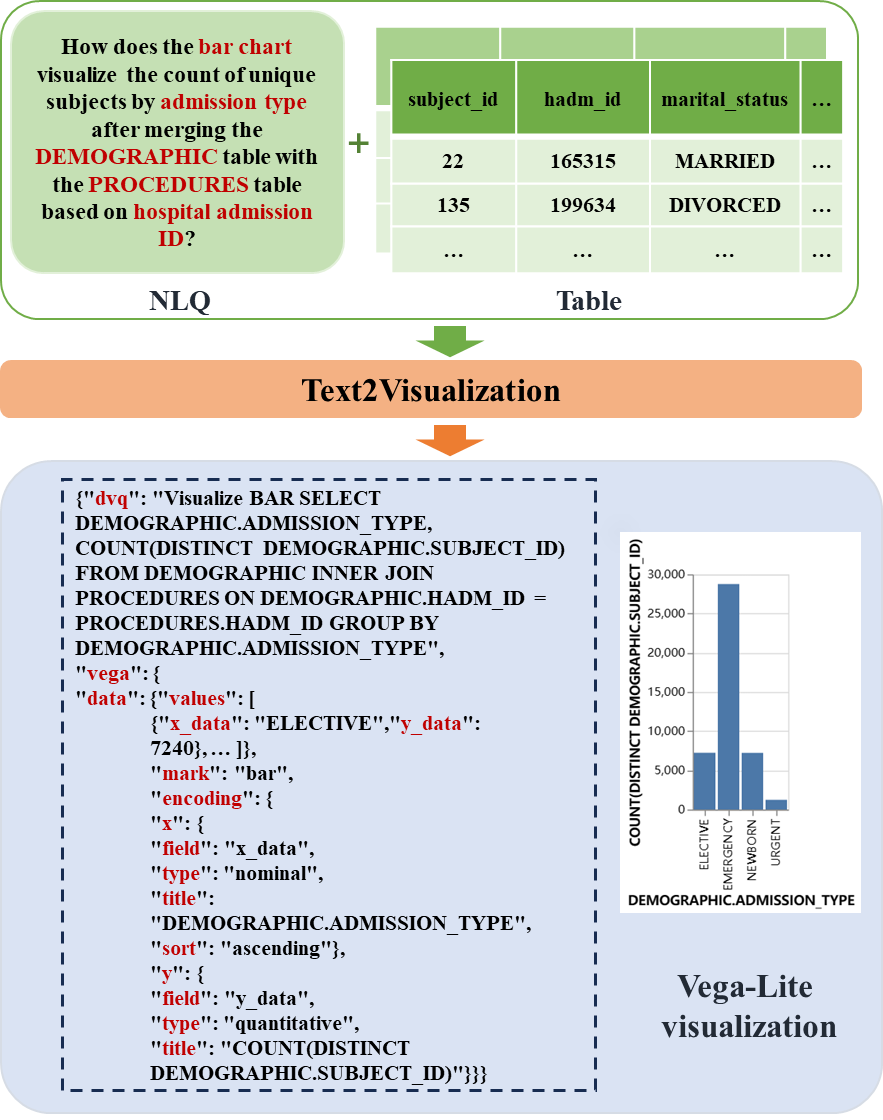}
\vspace{-5pt}
\caption{A text-to-visualization example from our proposed MedicalVis dataset, demonstrates how users can visually explore electronic medical record (EMR) data through natural language queries (NLQs).}
\vspace{-15pt}
\label{fig:example}
\end{figure}

One promising solution is applying data visualization techniques that transform EMR data into intuitive graphical summaries. Data visualization involves selecting data attributes, converting data, choosing visual coding methods, and rendering graphical representations. A crucial step in obtaining data visualization on a database involves writing visualization specifications in a declarative visualization language (DVL, e.g., Vega-Lite \cite{satyanarayan2016vega}), outlining the necessary data and its visual representation. However, proficiency in DVLs requires in-depth knowledge of domain-specific data and expertise in these languages, which is a significant barrier for beginners and non-technical users, and existing medical visualization tools are limited, with most focused on a single data type such as genomic sequences \cite{pandey2022genorec} and the domain of epidemiology\cite{crisan2021gevitrec}. And current text-to-visualization benchmarks focus on cross-domain applications, there remains a dearth of research on unified EMR visualization that accommodates the diverse content found in medical records. Although some scholars, such as Cai et al. \cite{Cai2023emrs}, have proposed a robust pre-training method based on weak supervision (WSRP) as a solution in the absence of large-scale labeled visualized medical data, there still exists a significant gap in the availability of effective visualization tools for researchers to efficiently explore and analyze EMR databases, posing a critical unresolved issue in the field. Furthermore, there is an urgent need to establish a benchmark to systematically compare the visualization techniques of EMRs. However, directly applying existing text-to-visualization methods on EMRs suffers two main challenges: (1) Medical terminology abbreviations. (2) Lack of large-scale healthcare text-to-visualization dataset.

To advance automatic data visualization technology in the medical field, we propose a novel data creation pipeline method that leverages LLMs (e.g. GPT-4 \cite{gpt4}) to develop a text-to-visualization benchmark tailored for EMRs. Fig.~\ref{fig:example} illustrates an example of text-to-visualization functionality provided within our MedicalVis dataset. In this instance, an NLQ and a database schema serve as inputs, requiring the text-to-visualization models to effectively comprehend the text, extract pertinent information for visualization, and ensure coherence and consistency among the resultant infographic elements. Specifically, the dataset construction pipeline includes three critical steps: 
(i) \textit{DVQs generation}\footnote{The data visualization utilizes an intermediate form known as data visualization query (DVQ) that could be executed to obtain the specifications and charts, following existing practice of general text-to-visualization studies \cite{song2022rgvisnet,luo2021synthesizing}.}.
By crafting appropriate SQL-to-DVQ prompts, we leverage LLMs to transform SQL statements from the EHR domain (e.g., MIMICSQL \cite{wang2020text}) into potential DVQs. 
(ii) \textit{DVQs filter}. 
The data visualizations generated from DVQs may contain inadequate or unsuitable components, such as pie charts with an excessive number of segments or queries that yield only single data points, rendering them unsuitable for effective visualization. To address these potential issues, we have implemented a comprehensive set of criteria to filter and refine the candidate DVQs based on their data visualizations. 
(iii) \textit{NLQs generation}. In the final step, through the design of a suitable DVQ-to-NLQ prompt, we employ LLMs as annotators to generate imperative and interrogative NLQs corresponding to each DVQ.

After the above-mentioned pipeline, we obtain the first large-scale text-to-visualization dataset for EMRs, entitled \textbf{MedicalVis}, comprising 35,374 examples. Following a thorough analysis and evaluation, we scrutinize the performance of various text-to-visualization models, namely Seq2Vis \cite{luo2021synthesizing}, ncNet \cite{luo2021natural}, and codeT5\cite{wang2021codet5}, utilizing the MedicalVis dataset. The experimental findings indicate the efficacy of the text-to-visualization approach in facilitating the visualization analysis of the EMR database. This not only unveils the information encapsulated within EMRs but also validates the dataset's high quality.

The contributions of this paper are summarized as follows:
\begin{itemize}
\item We release MedicalVis, the first large-scale benchmark for text-to-visualization in the EMR domain. This dataset will facilitate further research in automated data visualization and analysis of EMRs.

\item We introduce an innovative data creation pipeline that leverages LLMs to generate a large-scale, high-quality text-to-visualization dataset, thereby mitigating the need for labor-intensive manual annotation efforts.

\item We present MedCodeT5, a multi-task pre-trained model based on CodeT5, and extensively evaluate its performance against various text-to-visualization baselines. The experiment validates the feasibility of analyzing and visualizing EMR data through natural language interaction.
\end{itemize}

The rest of the paper is organized as follows. Section~\ref{sec:related} reviews the related work. 
Moving on to Section~\ref{sec:data}, we delve into the intricacies of the fundamental concept of the text-to-visualization task within the EMR domain. In section~\ref{sec:dataanal}, we introduce our newly proposed dataset, MedicalVis. 
Section~\ref{sec:methods} presents the details of the designed methods. 
Section~\ref{sec:exp} shows experimental results, followed by a conclusion in Section~\ref{sec:con}.

\section{Related work}
\label{sec:related}
We will review the related work from two fields: Medical Data Analysis and text-to-visualization in natural language processing(NLP).

\subsection{EMR Data Analysis}
EMRs and medical images represent critical components of medical big data, providing vast and diverse datasets. Eladio et al. \cite{Eladio24emr} have contributed by developing a medical dataset using the SQuAD V2.0 framework, which enriches the medical information available in EMR question answering. Recent advancements in deep learning have revolutionized the analysis and utilization of EMR data, leading to enhanced insights and applications in healthcare. For instance, Liang et al. \cite{bibm2014liang} addresse the limitations of traditional healthcare decision models by proposing a deep learning approach that integrates feature representation and learning. Using a modified convolutional deep belief network demonstrates improved performance over shallow models in predicting outcomes for hypertension and Chinese medical diagnoses. Gao et al. \cite{Gao21} presented SMP-Graph, a novel method for unsupervised semantic graph representation enhanced with structural information, designed to boost the efficiency of medical procedure coding in EMRs. This approach integrates contextualized information within a graph framework, facilitating more accurate and detailed analysis of medical data.
Jia et al. \cite{bibm2017jia} introduce a spatiotemporal autoencoder (STAE) designed to learn features from large-scale patient data, which often includes spatial and temporal dimensions and may have missing observations. STAE effectively identifies patterns and dependencies in patient data, enabling improved classification through compact patient representations.
Another example is GenoREC \cite{pandey2022genorec}, which recommends interactive genomic data visualizations based on specific data and task requirements, empowering analysts to explore and apply recommended visualizations effectively.

Additionally, text-to-SQL methods are increasingly used in natural language processing to manage EMR data \cite{wang2020text, lee2022ehrsql}. These methods focus on extracting relevant data for specific queries but may not address comprehensive data analysis needs. In contrast, text-to-visualization approaches are tailored to support data analysis tasks, allowing users to uncover critical insights without grappling with technical details like data formats or visualization methods.

Furthermore, text-to-visualization shows promise in enabling non-expert users to create data visualizations. Zhang et al. \cite{healthlens} introduce HealthLens, the first user-friendly visualization tool in the EMR domain that eliminates the need for prior knowledge of DVLs. Given the richness of information in EMRs, there is a need for further development and specialization of text-to-visualization methods and datasets to enhance efficiency in patient care and advance clinical research.

\subsection{text-to-visualization in NLP}
Research in text-to-visualization has spanned decades. Initially, text-to-visualization systems relied on symbol-based natural language process approaches, employing heuristics, rule-based architectures, and probabilistic grammar-based methods to interpret queries. 

Flowsense \cite{yu2019flowsense} further advanced text-to-visualization by integrating state-of-the-art semantic parsing techniques within data flow systems, enabling users to employ queries for various data flow graph editing tasks. Notably, it offers users explicit insights into identified datasets and chart-specific instructions to enhance data-flow context awareness. 
NL4DV \cite{narechania2020nl4dv}, a Python toolkit, follows suit, providing a parsed specification model in JSON format upon receiving tabular datasets and corresponding queries. This toolkit also includes Vega-Lite specification lists and analysis properties associated with the input query.
However, these approaches are constrained by rule-based methods, limiting their adaptability to diverse user inputs and impeding support for free-form queries. 
Recent advancements have turned to deep learning NLP techniques, like language representation models, to address these limitations. ADViSor \cite{liu2021advisor}, for instance, introduces a deep learning-based pipeline for generating visualization strategies in response to natural language questions regarding tabular data. The ADViSor process comprises two primary steps: NLQ to SQL data extraction and rule-based visual generation. 
The availability of the cross-domain NvBench\cite{luo2021synthesizing} dataset since 2021 has facilitated extensive neural network model training with numerous (NLQ, Visualization) pairs. ncNet \cite{luo2021natural} employs Transformer to translate queries into visual outcomes. 
RGVisNet \cite{song2022rgvisnet} innovates with a hybrid retrieval-generation framework, utilizing a search and generation method. %It retrieves relevant candidate data visualization queries from a code base, modifying them to generate desired data visualization queries. 
Prompt4Vis \cite{li2024prompt4vis} introduces contextual learning into the text-to-visualization task, integrating a multi-target example mining module to identify genuinely effective examples.

Despite these previous efforts, there has been a lack of work on visualization on electronic medical domain. Our work is the first to address this gap by constructing the benchmark dataset for medical data visualization. We validate the performance of numerous DNN-based and LLM-based approaches on this dataset, thereby advancing the field of medical data visualization. Through our pioneering contributions, we pave the way for further developments in eliciting insights from EMRs data through effective visual representations.

\section{Dataset Construction}
\label{sec:data}

In this section, we introduce our proposed framework that leverages the capabilities of LLMs to enable the automatic generation of data for visualizing EMRs. This method exploits the strengths of LLMs to extract relevant information from EMRs, which is subsequently utilized for medical data visualization.

\begin{figure*}[ht]
  \centering
  \includegraphics[width=1\linewidth]{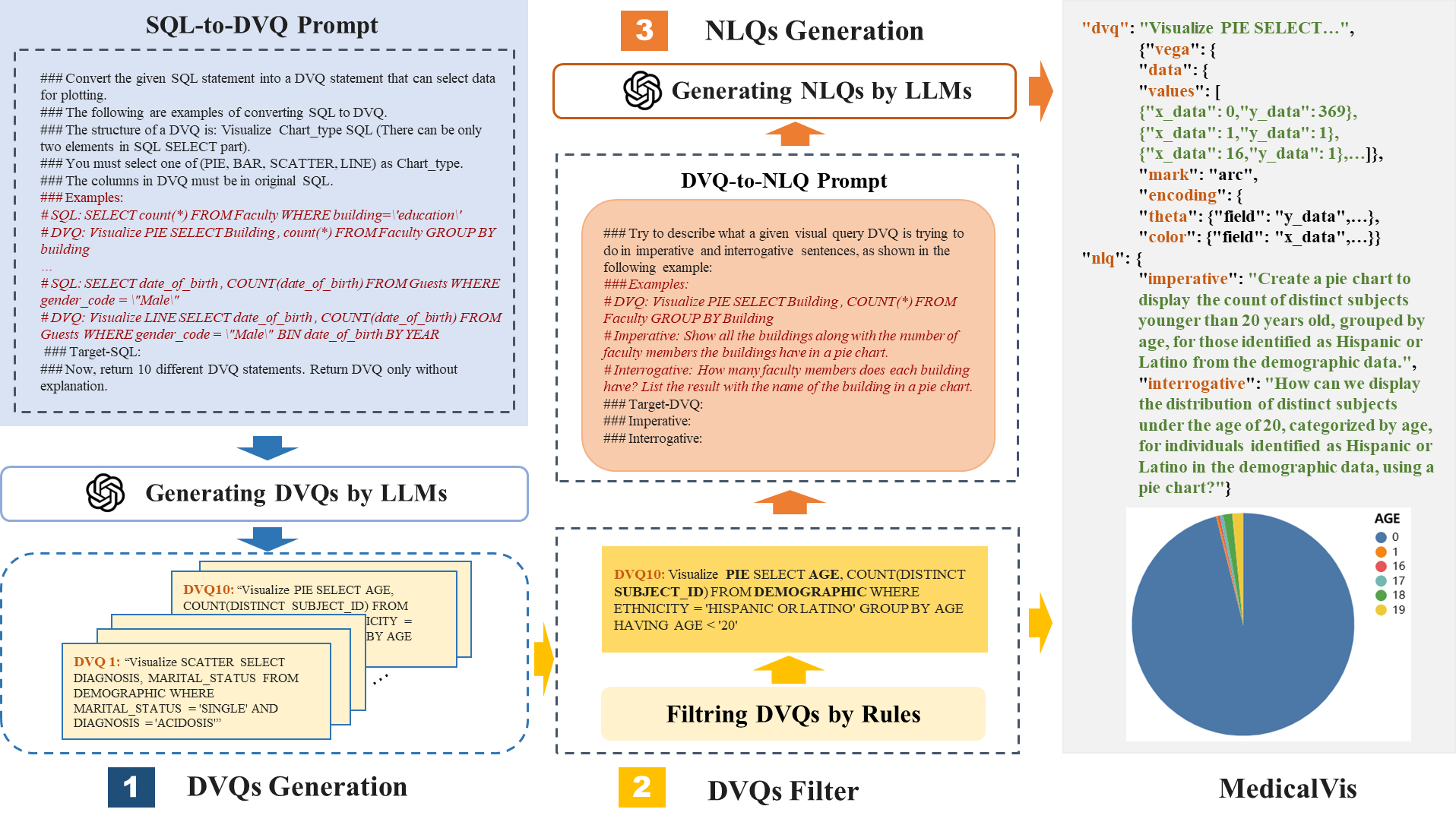}
  \caption{The proposed pipeline for creating an EMR text-to-visualization benchmark dataset. The pipeline comprises three key stages: (i) \textit{DVQs Generation} from EMR data, (ii) \textit{DVQs Filter} to check quality and ensure high-quality visualizations, and (iii) \textit{NLQs Generation} to translate from data visualizations to textual descriptions by LLMs as annotators.}
  \label{fig:framework}
  \vspace{-15pt}
\end{figure*}

\subsection{The Overall Pipeline}

The overall data generation pipeline, as shown in Fig.~\ref{fig:framework}, consists of the following three main stages. 
The first stage is DVQ candidate generation, where we use prompt techniques from LLMs to translate structured SQL queries into understandable and analyzable DVQs. 
The next stage is DVQ filtering and checking, where we employ a rule-based filter to screen the candidate visualization queries. This step involves filtering out overly complex, cumbersome, and unreasonable DVQ statements, ensuring that only high-quality and informative visualization results are retained. Through this process, potentially misleading or incomplete visualization queries generated by the model are eliminated. The final stage is NLQ generation, which focuses on transforming the generated data visualization results into understandable natural language questions. In this stage, we manually align the generated natural language with the subset of filtered DVQ and conduct matching tests to ensure that the generated natural language is fluent and the syntax directing to the DVQs is entirely correct. In the following sections, we will provide detailed descriptions and explanations for each of above steps.

\subsection{DVQ Candidate Generation}
The use of LLMs provides a convenient way for us to transform SQL queries. We use prompt methods to convert SQL queries into DVQs, as depicted in Step 1 of Fig.~\ref{fig:framework} The initial transformation of an SQL query $Q_{sql}$ into a visualization query $Q_{dvq}$ can be formally described as follows: an SQL query $Q_{sql}$ is provided as input to the model, using LLMs such as GPT-4, the query $ Q_{sql}$ is parsed and restructured into a visualization query $Q_{dvq}$. This process can be represented as:
\begin{equation}
  Q_{dvq} = \text{LLM}(Q_{sql}; P_{S2D})
\end{equation}
where $ \text{LLM}$ denotes the Large Language Model that interprets and converts the SQL query and $P_{S2D}$ denotes our appropriately crafted SQL-to-DVQ prompt. The resulting $ Q_{dvq}$ provides the instructions for generating the desired data visualization.

To achieve this, the instructional prompt $P_{S2D}$ is utilized, as shown in the SQL-to-DVQ Prompt example in Fig.~\ref{fig:framework}. The prompt guides the model on how to transform SQL queries into DVQs, with the aim of preserving the fundamental elements of the SQL query while reformatting it into a DVQ suitable for visualization purposes. For instance, we convert SQL statements like :
\textit{`SELECT \textbf{``DISCHTIME''}, count(``DAYS\_STAY'') FROM \textbf{DEMOGRAPHIC} WHERE ``SUBJECT\_ID'' = ``9575'' GROUP BY ``DISCHTIME'' '} into DVQ statements like \textit{`\textbf{Visualize SCATTER} SELECT \textbf{``DISCHTIME''}, count(``DAYS\_STAY'') FROM \textbf{DEMOGRAPHIC} WHERE ``SUBJECT\_ID'' = ``9575'' GROUP BY ``DISCHTIME'' \textbf{BIN ``DISCHTIME'' BY MONTH}`}. In this example, the SQL query $ Q_{sql}$ is to count the number of "DAYS STAY" corresponding to DISCHTIME. The LLM transforms it into a visualization query $ Q_{dvq}$, specifically a scatter chart, and additional pairs of months to group DISCHTIME.

\subsection{DVQs Checking and Filtering}
We utilize a rule-based filter $R(\cdot)$ to eliminate undesirable candidate DVQs based on their data visualizations. When given a DVQ $q$, $R(q)$ will output either true (indicating a DVQ with good visualization) or false (indicating a DVQ with bad visualization). Table \ref{table1} presents a set of rules for inspecting and filtering DVQs, which are intended to detect and remove unsatisfactory visualizations from the candidate DVQ set $T_Q$. By applying the rule-based filter to each DVQ, those that meet the conditions specified in the table will be pruned from $T_Q$. 

The rules in Table \ref{table1} are categorized based on chart type. For all chart types, a DVQ will be eliminated if it meets certain criteria, such as retrieving an excessive number of values ($N > 10000$ or $N = 1$) or having nominal y-axis values. Additionally, line and scatter charts will be removed if they retrieve exactly two values and have nominal x-axis values. Pie charts will be removed if they retrieve more than ten values, and bar charts will be removed if they retrieve more than twenty-six values. By adhering to these rules, we can effectively eliminate unsatisfactory visualizations from the candidate visualization set $T_Q$ and obtain a refined set of satisfactory visualizations denoted as $T_Q'$.

\begin{table}[htb]
\small
\centering
\caption{Rules for DVQs checking and Filtering}
$N$: Number of values the query retrieves
\label{table1}
\begin{tabular}{lc}
\noalign{\smallskip}
\toprule
\noalign{\smallskip}
\textbf{Type of the chart} & \textbf{Rule} \\
\toprule
\multirow{3}{8em}{All} 
& $N > 10000 \vee N = 1$,\\
&$v$ with nomianl y-axis values \\
\noalign{\smallskip}
\hline

\noalign{\smallskip}
\multirow{2}{8em}{Line/Scatter} 
&$N = 2$, \\
&$v$ with nomianl x-axis values\\
\hline

\noalign{\smallskip}
\multirow{1}{8em}{Pie} 
&$N > 10$ \\
\noalign{\smallskip}
\hline

\noalign{\smallskip}
\multirow{1}{8em}{Bar} 
&$N > 26$ \\
\bottomrule
\end{tabular}
\vspace{-10pt}
\end{table}

\begin{figure*}[t!]
 \begin{minipage}[b]{.5\linewidth}
  \centering
  \includegraphics[width=0.7\columnwidth]{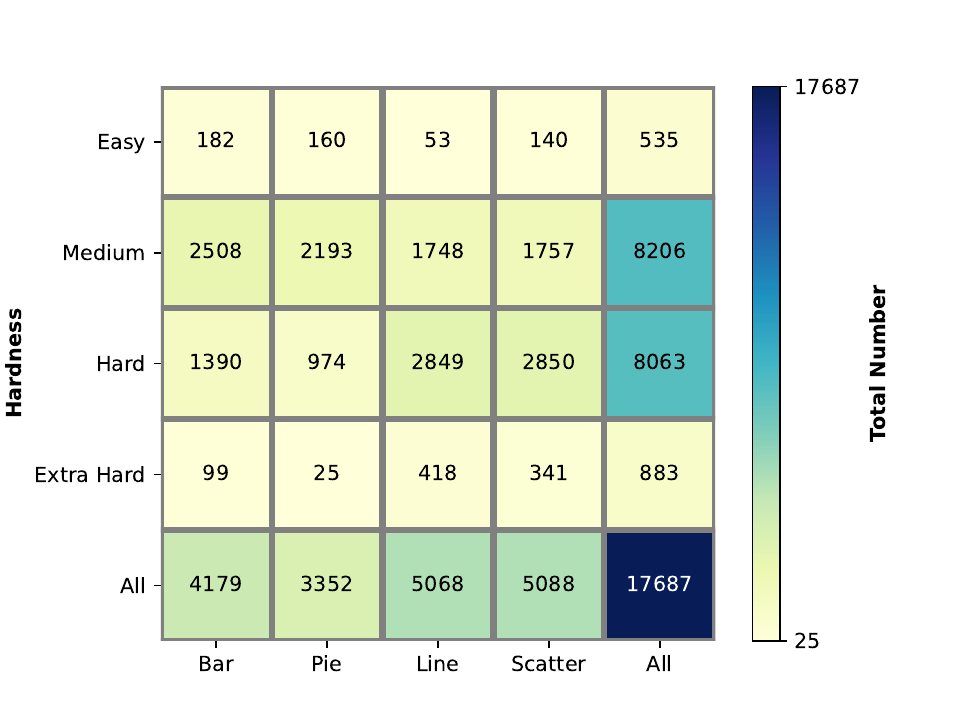}
  \caption{Visualization types vs. hardness}
 \end{minipage}\hfill
 \begin{minipage}[b]{.5\linewidth}
  \centering
    \includegraphics[width=\columnwidth]{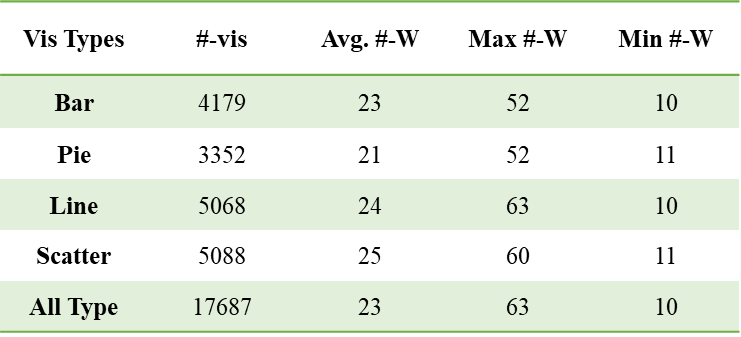}
  \captionof{table}{NLQ Types and Word Counts (\#-W) in MedicalVis}
  \label{tablesta}
 \end{minipage}
 \vspace{-20pt}
\end{figure*}

\subsection{NLQs Generation}

Visualizations such as charts and graphs visually encode complex information. To generate explanatory text from them, a more systematic analysis of the different elements of the visualization is required in order to explain their representations.
Firstly, it is necessary to determine the main theme or purpose of the visualization by examining the title, axis labels, and other textual components corresponding to the diagram.
Subsequently, breaking down various visual components like bar charts, line charts, pie charts, etc., and analyzing their relationships with each other and the scales used becomes essential. For instance, in a bar chart, a higher bar indicates a higher value. Trends and patterns are identified through changes in values over time within visual representations. These observations serve as foundations for generating sentences and paragraphs that elucidate key points within visualizations. Describing patterns, trends, and insights gained using precise language to refer to specific data points and comparison values is crucial.

The ultimate goal is to articulate complex visual displays using clear language. The DVQ-to-NLQ prompt design, as shown in Fig.~\ref{fig:framework}, aims at making natural language design more aligned with visual effects while being compatible with human usage habits.

\section{MedicalVis Dataset Analysis}
\label{sec:dataanal}

\subsection{Overall Statistics}
MedicalVis dataset includes 35,374 (NLQ, DVQ) pairs, and basic statistics are provided in Table \ref{table4}. The dataset consists of DVQs distributed over 5 tables with a total of 49 columns. This variation allows researchers to explore different levels of data granularity. On average, each DVQ has 1.10 aggregated columns, and includes 1.41 conditions.  This enables comprehensive summary and analysis operations, including calculations for averages, sums, and other statistical measures essential for medical analytics.

\begin{table}[htb]
\small
\centering
\caption{Statistics of MedicalVis dataset.}
\label{table4}
\renewcommand{\arraystretch}{1.5}
% Increase row spacing
\begin{tabular}{lc}

\noalign{\smallskip}
\toprule
\textbf{Data} & \textbf{Value} \\
\toprule

\# of tables involved & 5 \\
\hline
\# of columns involved & 49 \\
\hline
\# Average DVQ length & 21 \\
\hline
\# Average aggregation columns & 1.10 \\
\hline
\# Average conditions & 1.41 \\
\bottomrule
\end{tabular}
\vspace{-10pt}
\end{table}

\subsection{Detailed Analysis}
The MedicalVis dataset consists of 35,374 (NLQ, DVQ) pairs and 17,687 distinct visualizations. In the following sections, we will discuss the dataset in more detail.

\subsubsection{Visualization types}
MedicalVis includes 17,687 distinct visualizations across four types: bar (histogram), pie, line, and scatter. Scatter charts are the most prevalent, making up nearly 29\% of the total, while pie charts are the least common. Scatter plots play a crucial role in healthcare data visualization, which can be pivotal for correlation analysis and trend observation, enabling healthcare professionals to visualize and understand relationships between variables and helping to identify outliers or unusual data points.

\subsubsection{Visualization hardness}
Intuitively, not all NLQ2DVQ tasks are equally difficult. Similar to nvBench, we categorize DVQs into four levels: easy, medium, hard, and extra hard.

To evaluate the complexity of DVQs, we introduce a scoring framework that categorizes DVQs into different levels based on their structural components. Each component contributes to the overall complexity score \( S_{\text{total}} \), which is calculated by aggregating scores from the following clauses: \( S_{\text{select}} \) (number of aggregators and distinct columns in the SELECT part), \( S_{\text{table}} \) (number of JOIN operations in the FROM part), \( S_{\text{group}} \) (number of  clauses in the GROUP BY part), \( S_{\text{order}} \) (number of clauses in the ORDER BY part), \( S_{\text{conditions}} \) (number of conditions in the WHERE, HAVING and LIMIT part), and \( S_{\text{bin}} \) (1 if there is a BIN BY part in DVQ, otherwise 0).

S$_{total}$ is computed by summing up the scores of all the clause we mentioned about. The difficulty levels of the questions can be categorized as: Easy ($S_{total}\leq2$)
, Medium ($2<S_{total}\leq4$), Hard ($4<S_{total}\leq6$), and Extra Hard ($S_{total}>6$).

\begin{figure*}[t]
  \centering
  \includegraphics[width=1\linewidth]{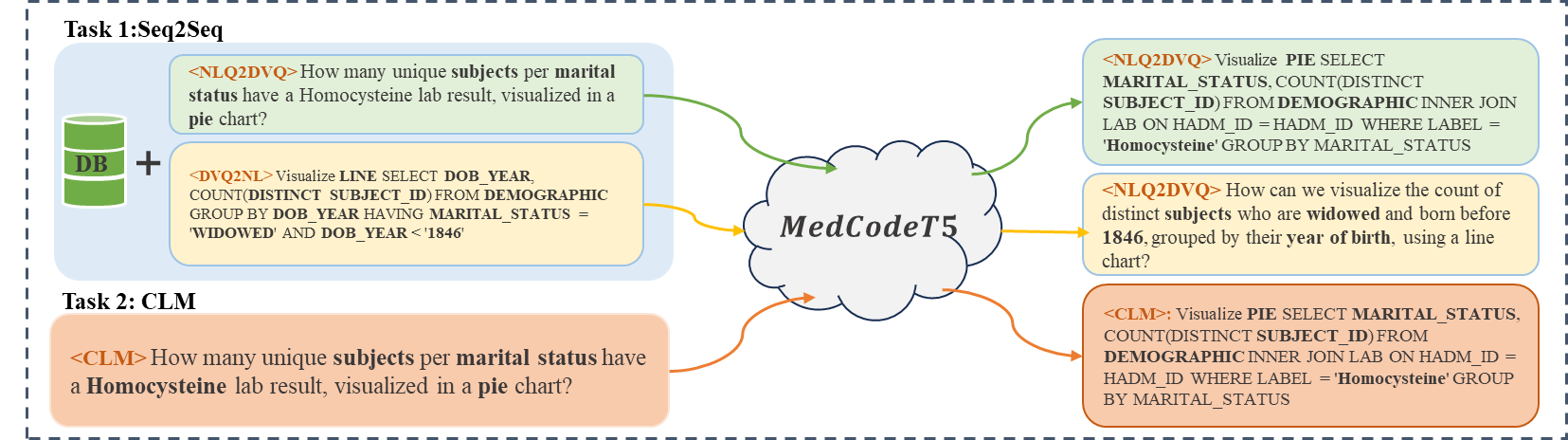}
  \caption{
  Illustration of the mixed-task pretraining approach for the MedCodeT5 model. The approach includes (i) converting natural language queries to data visualization queries (NLQ2DVQ), (ii) converting data visualization queries to natural language (DVQ2NLQ), and (iii) sequence-to-sequence causal language modeling (CLM).}
  \label{fig:medt5}
  \vspace{-5pt}
\end{figure*}

\subsubsection{NLQs}
Table \ref{tablesta} displays the number of DVQs of different visualization types and the length of NLQs of the corresponding type. On average, each NLQ consists of approximately 23 words. Some NLQs have higher word counts due to the complexity of the visual query, which may involve filtering and join operations. At the same time, aiming at the medical domain, to make the expression of NLQ unambiguous, this usually also makes our NLQS length generally longer.

\section{LLM-based text-to-visualization Methods}
\label{sec:methods}
In this section, we explore the feasibility of generating visualizations from natural language descriptions in EMRs. As illustrated in Fig.~\ref{fig:medt5}, we conducted a mixed-task pretraining of the MedCodeT5 model, incorporating three key tasks: including NLQ2DVQ, DVQ2NLQ, and sequence-to-sequence CLM. This comprehensive pretraining approach enables the model to learn meaningful context representations and effectively recover missing information within complex DVQs.

\subsection{Our Proposed Method: MedCodeT5}

Given the widespread adoption of LLMs in NLP tasks, the CodeT5 model developed by Wang et al\cite{wang2021codet5}. is specifically designed for code-related tasks and offers great potential for improving our results. To enhance its performance in medical visualization, we optimized the CodeT5 model using the MedicalVis dataset and trained a version dedicated to medical DVQ and NLQ generation, named MedCodeT5. During the training of MedCodeT5, we specifically trained the CodeT5 model on sequence-to-sequence and CLM tasks to enhance its language generation performance. This mixed-task pretraining approach included converting natural language queries to data visualization queries (NLQ2DVQ) and vice versa (DVQ2NLQ). 
Furthermore, we employed sequence-to-sequence CLM to enhance the model's ability to generate coherent and contextually accurate language. Referring to the method of CodeT5+\cite{wang2023codet5plus}, we randomly select a pivot position and treat NLQ as the source sequence and subsequent DVQ as the target output. To achieve the sequence-to-sequence CLM objective, we restrict pivot positions to between 10\% and 80\% of the entire sequence for uniform sampling, while adding a special token [CLM] at the beginning of the entire sequence. The sequence-to-sequence training involves mapping input sequences to output sequences using Transformer architecture's encoder-decoder structure, whereas CLM focuses on predicting the next token in a sequence to improve fluent text generation. By leveraging these tasks, we robustly trained MedCodeT5 to handle the unique requirements of medical data visualization and NLQ generation, resulting in enhanced accuracy, and robust language generation. This optimization allowed MedCodeT5 to deliver high performance tailored to the specific requirements of medical data visualization and NLQ generation.

\section{Experiments and Analysis}
\label{sec:exp}

In this section, we describe our experimental setup and present the evaluation results. Through these experimental settings and evaluation processes, we validate the effectiveness and feasibility of the MedicalVis dataset in addressing complex text-to-visualization tasks.

\subsection{Experimental Setup}

\subsubsection{Dataset} According to the segmentation of the MIMICSQL dataset, we performed a similar partitioning of our MedicalVis dataset, dividing it into three different subsets: a training set, a validation set, and a test set. The training set consists of 8000 samples, and the validation set and test set each consist of 1000 samples.

\subsubsection{Text-to-visualization Methods} We choose the following methods since they are the most popular ones in text-to-visualization fields. 
\begin{itemize}
\item\textit{Seq2Vis\cite{luo2021synthesizing}}: Seq2Vis is a widely used model that utilizes a standard encoder-decoder architecture with an attention mechanism for enhanced performance.

\item\textit{ncNet\cite{luo2021natural}}: ncNet is a Transformer-based model that allows users to select chart templates as additional input. This enables the model to incorporate information from the chosen chart templates into the natural language processing.

\item\textit{CodeT5\cite{wang2021codet5}}: CodeT5 is a Transformer-based model designed for code generation and understanding, leveraging multi-task pre-training to excel in tasks such as code completion and translation.

\item\textit{MedCodeT5}: MedCodeT5 excels in converting NLQ to DVQ and vice versa, facilitating seamless interaction with medical databases. Its robust pretraining on diverse tasks ensures high performance in understanding and generating medical data visualizations.

\end{itemize}

\subsubsection{Evaluation Metrics} In order to better explore the performance of our dataset across different models, we have employed exact match accuracy as a measure for the components used in visualizations, which include visualization matching accuracy, data value matching accuracy, axis matching accuracy, and overall matching accuracy.

\begin{equation}
\text{$M_\text{vis}$} = \frac{\text{$V_\text{correct}$}}{\text{$N_\text{total}$}},
\text{$M_\text{data}$} = \frac{\text{$D_\text{correct}$}}{\text{$N_\text{total}$}},
\text{$M_\text{axis}$} = \frac{\text{$A_\text{correct}$}}{\text{$N_\text{total}$}}
\end{equation}

Where $M$ means match accuracy, $V$,$D$ and $A$ represent visualization chart type, data, horizontal and vertical axes. {$N_\text{total}$} indicates the total number of data items, that is, the total number of DVQs in the dataset or the overall data items after Vega-Lite transfer. The overall matching accuracy calculation follows the same procedure, where we determine the overall matching degree by assessing the exact matching between the predicted DVQ and the ground truth. It is worth mentioning that we are using a strict matching metric, even if the data value in the SQL part of the DVQ is wrong, or if the column name A and column name B appear in reverse order, we will determine a mismatch as a whole.

\subsection{Experimental Analysis}

\subsubsection{Performance Comparison}
We conducted a comparative analysis of various benchmarks for large-scale language models and deep learning generative models using our datasets, including Seq2Vis, ncNet, CodeT5, and MedCodeT5. For training large-scale language models, we utilized data fine-tuning to leverage the capabilities of pre-trained models on extensive datasets. Additionally, we incorporated self-consistency techniques when employing LLM for few-shot learning to improve results. We queried the CodeT5 output multiple times ($N=3$ in our experiments) and selected the answer with the highest probability. By combining data fine-tuning and self-consistency techniques, we fully exploited the advantages of large models while overcoming potential instability in few-shot learning to achieve more accurate and reliable predictions. 

Table \ref{tablefour} illustrates the performance of various methods in generating data visualization queries. MedCodeT5 stands out with the highest overall accuracy at 60.4\% and excellent axis accuracy of 97.2\%. Both CodeT5 and MedCodeT5 achieve high data accuracy of 97.2\% and perfect visualization type accuracy at 100\%. MedCodeT5's superior overall performance likely stems from its use of multitask learning, which enhances its ability to handle diverse data and improves generalization across tasks.

\begin{table}[htb]
\small
\centering
\caption{Matching Accuracy(\%)}
\label{tablefour}
\begin{tabular}{lcccc}
\noalign{\smallskip}
\hline
\noalign{\smallskip}
Method & Data & Axis & Vis\_type & Overall\\
\toprule

Seq2Vis
&33.6  &22.6  &71.0 &8.7\\
ncNet
&85.9   &-   &100.0   &55.0 \\ 
CodeT5
&97.2   &90.3  &100.0  &46.0 \\

\textbf{MedCodeT5}
&\textbf{97.2}  &\textbf{97.2}  &\textbf{100.0} &\textbf{60.4}\\
\bottomrule
\end{tabular}
\vspace{-10pt}
\end{table}

\subsubsection{Detailed Analysis}
To compare the performance differences between models in more detail, we compared the hardness of the dataset. In the aforementioned dataset analysis, we categorized its hardness into four levels based on the composition structure of DVQs: easy, medium, hard, and extremely hard. As the hardness level escalates, the composition structure becomes more intricate, entailing increasingly complex clauses and nested relationships.

The performance of the four methods on problems with varying levels of difficulty is compared in Table \ref{tabhardness}, including Seq2Vis, ncNet, CodeT5, and MedCodeT5. The analysis results indicate that MedCodeT5 demonstrates superior performance across all difficulty levels, achieving an overall score of 60.4\% and notably scoring 21.8\% on extremely difficult problems, showcasing its robust adaptability and effectiveness. ncNet also exhibits strong performance on medium and challenging problems, with an overall score of 55.0\%. In contrast, CodeT5 displays consistent efficiency across difficulty levels but achieves a lower overall score of 46.0\%. On the other hand, Seq2Vis shows poor performance on hard and extremely hard problems, obtaining an overall score of only 8.7\%, indicating its limited capability to handle complex problems.

\begin{table}[th!]
%\small
\centering
\caption{Hardness Comparison(\%)}
\scalebox{0.87}{
\begin{tabular}{lccccc}
\noalign{\smallskip}
\hline
\noalign{\smallskip}
Method & Easy & Medium & Hard & Ex. Hard & Overall \\
\toprule

Seq2Vis
&33.3  &14.9  &1.86 &0 &8.7\\
ncNet
&72.0   &58.9   &47.8  &12.1 &55.0 \\ 
CodeT5
&50.0  &51.5  &45.4  &11.9 &46.0\\

\textbf{MedCodeT5}
&\textbf{75.4}  &\textbf{66.4}  &\textbf{59.1}  &\textbf{21.8} &\textbf{60.4}\\

\bottomrule
\label{tabhardness}
\end{tabular}
}
\vspace{-5pt}
\end{table}

\subsubsection{Case Study} In this section, we demonstrate the practical application of our EMR visualization dataset and compare the performance of MedCodeT5 with other baselines through a concrete case study. 
Given the NLQ: "Create a scatter plot to display the count of hospital admissions for each age group under 80 with a history of venous thrombosis or embolisms", our DVQ evaluation focuses on several key aspects. Firstly, it is essential to identify patients under the age of 80 with a history of venous thrombosis or embolism. Subsequently, the generated DVQ must meticulously summarize and analyze admission records for each age group to ensure accuracy. Finally, the data can be transformed into a scatter plot to visually present the key findings and research results clearly and intuitively.

Seq2Vis manages to complete the basic parts of the DVQ but generates an incomplete version, missing one condition and the GROUP BY clause, and fails to capture the full complexity of the query. Meanwhile, ncNet incorrectly handles column and condition selection. Although CodeT5 nearly generates a correct DVQ, it mistakenly selects the wrong condition column. In contrast, MedCodeT5 can more reasonably generate the corresponding DVQ according to the NLQ input.

\begin{table*}[th!]
  \centering
  \includegraphics[width=0.98\linewidth]{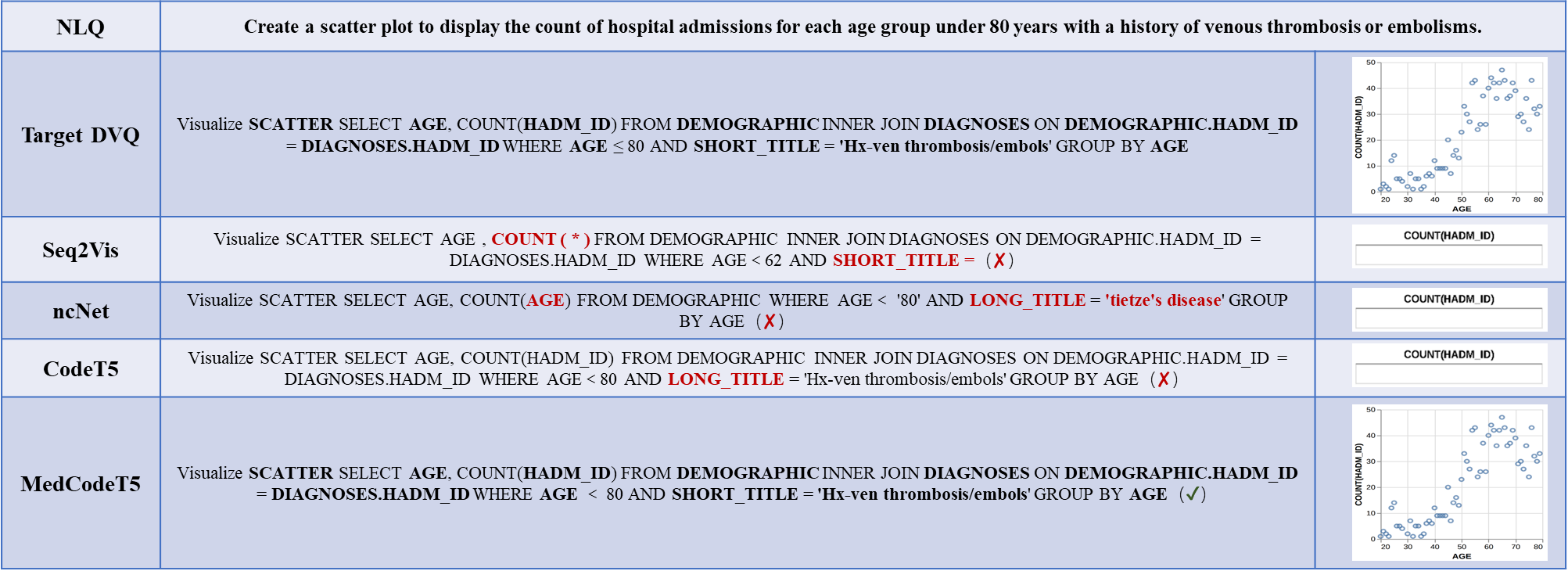}
  \caption{Case Study: DVQs generated by diverse text-to-visualization methods on MedicalVis, accompanied by their corresponding charts (Errors indicated in \textcolor{red}{red}).}
  \label{tab:case_study}
  \vspace{-15pt}
\end{table*}

\section{Conclusion}
\label{sec:con}
We propose a novel pipeline to create a text-to-visualization benchmark dataset tailored for EMRs. This dataset enables users to generate visualizations from NLQs about EMR data. It consists of paired medical records, NLQs, and corresponding visualizations. To alleviate manual annotation efforts, we leverage LLMs for data creation. We evaluate state-of-the-art text-to-visualization techniques and LLM-based approaches on these new datasets, confirming the feasibility of generating EMR visualizations from NLQs. Our work facilitates standardized assessment of EMR visualization methods, advancing this impactful application field and promoting the elicitation of medical insights through visualization.

\linespread{1.0}
\bibliographystyle{IEEEtran}
% \bibliography{ref-short}
% Generated by IEEEtran.bst, version: 1.14 (2015/08/26)

\end{document}